# Convolutional Neural Network Architecture for Recovering Watermark Synchronization

Wook-Hyung Kim, Jong-Uk Hou, Seung-Min Mun, and Heung-Kyu Lee

*Abstract*—**Digital content has always been subject to copyright infringement due to its ease of duplication. In recent years, the amount of real-time content provided by Internet browsers and smartphone applications has been increasing due to the rapid development of network environments. However, since these real-time contents can be captured and downloaded very easily, copyright infringement has become a serious problem. In order to reduce the loss caused by copyright infringement, copyright owners insert a watermark in the content to protect the copyright using illegal distribution route tracking and copyright authentication. However, whereas many existing watermarking techniques are robust to signal distortion such as compression, they are vulnerable to geometric distortion that causes synchronization errors. In particular, capturing real-time content in Internet browsers and smartphone applications is problematic because geometric distortion such as scaling and translation frequently occurs. In this paper, we propose a convolutional neural network-based template architecture that compensates for the disadvantages of existing watermarking techniques that are vulnerable to geometric distortion. The proposed template consists of a template generation network, a template extraction network, and a template matching network. The template generation network generates a template in the form of noise and the template is inserted into certain pre-defined spatial locations of the image. The extraction network detects spatial locations where the template is inserted in the image. Finally, the template matching network estimates the parameters of the geometric distortion by comparing the shape of spatial locations where the template was inserted with the locations where the template was detected. It is possible to recover an image in its original geometrical form using the estimated parameters, and as a result, watermarks applied using existing watermarking techniques that are vulnerable to geometric distortion can be decoded normally.**

*Index Terms*—**Digital watermark, Deep-learning watermark, Depth-image-based rendering (DIBR) watermark, Copyright protection, Template watermark**

## I. INTRODUCTION

Digital content has always been subject to copyright infringement due to its ease of duplication. Recently, as the market for real-time content services such as web-comics (also called webtoons) and video streaming services such as YouTube has grown very rapidly, the problem of copyright infringement is increasing. These services, unlike traditional paid services, offer free content and earn revenue by advertising to users. As a result, these services are more vulnerable to illegal copying because they are easier to access than services that are provided for a fee. In addition, these services are provided on Internet browsers or as smartphone applications, making it easier to replicate the content with the Internet browsers' or applications' downloading or capturing functions.

Watermarking techniques have emerged to reduce the loss caused by piracy. These techniques minimize loss through the insertion of invisible information in the content to enable tracking of illegal distribution routes and copyright authentication. Recently, Web-comic companies and content providers of IPTV have started actively using these techniques to track and punish illegal distribution.

To date, numerous watermarking techniques have been proposed to protect the copyright of content. Methods of inserting watermarks into various transform domains such as discrete cosine transform (DCT), discrete Fourier transform (DFT), radon transform, curvelet transform, Dual-tree complex wavelet transform (DT-CWT), and contourlet transform have been proposed [1]–[6]. Various insertion methods such as spread-spectrum, quantization index modulation (QIM), angle QIM, and absolute angle QIM have also been proposed [7]–[10]. However, these watermarking techniques often show weaknesses after geometric distortion. Because geometric distortion causes synchronization errors, it is not easy to ensure robustness of these watermarks to geometric attacks.

In order to reduce the vulnerability to geometric attacks, many studies have been carried out to identify the transform domain and insertion methods with invariant characteristics against geometric distortions [11]–[14]. However, these algorithms have drawbacks such as low invisibility, lack of bit capacity such as zero-bit watermarking, vulnerability to specific geometric distortion such as translation, or necessity of additional information.

Many template-based watermarking methods for decoding watermarks using template matching techniques have been proposed [15]–[18], but these methods also have a drawback in that it is difficult to insert sufficient copyright information because the bit capacity is low. In addition, since these templates are additionally inserted over the watermarked image in which the copyright information is embedded, the invisibility is further reduced, and the template and the watermark signal may interfere with each other.

These watermarking techniques robust to geometric attacks have many disadvantages compared to watermarking techniques that do not consider geometric attacks. In addition, if the synchronization problem is solved from the geometrically distorted image, the watermark can be normally decoded. For this reason, watermarking techniques that do not consider geometric attacks are used in many cases. If a geometric attack occurs, watermark decoding is performed after the image is recovered into the geometric characteristics of the original

W.-H. Kim is currently working toward his Ph.D. degree in Multimedia Computing Lab., School of Computing, KAIST, e-mail: whkim@mmc.kaist.ac.kr

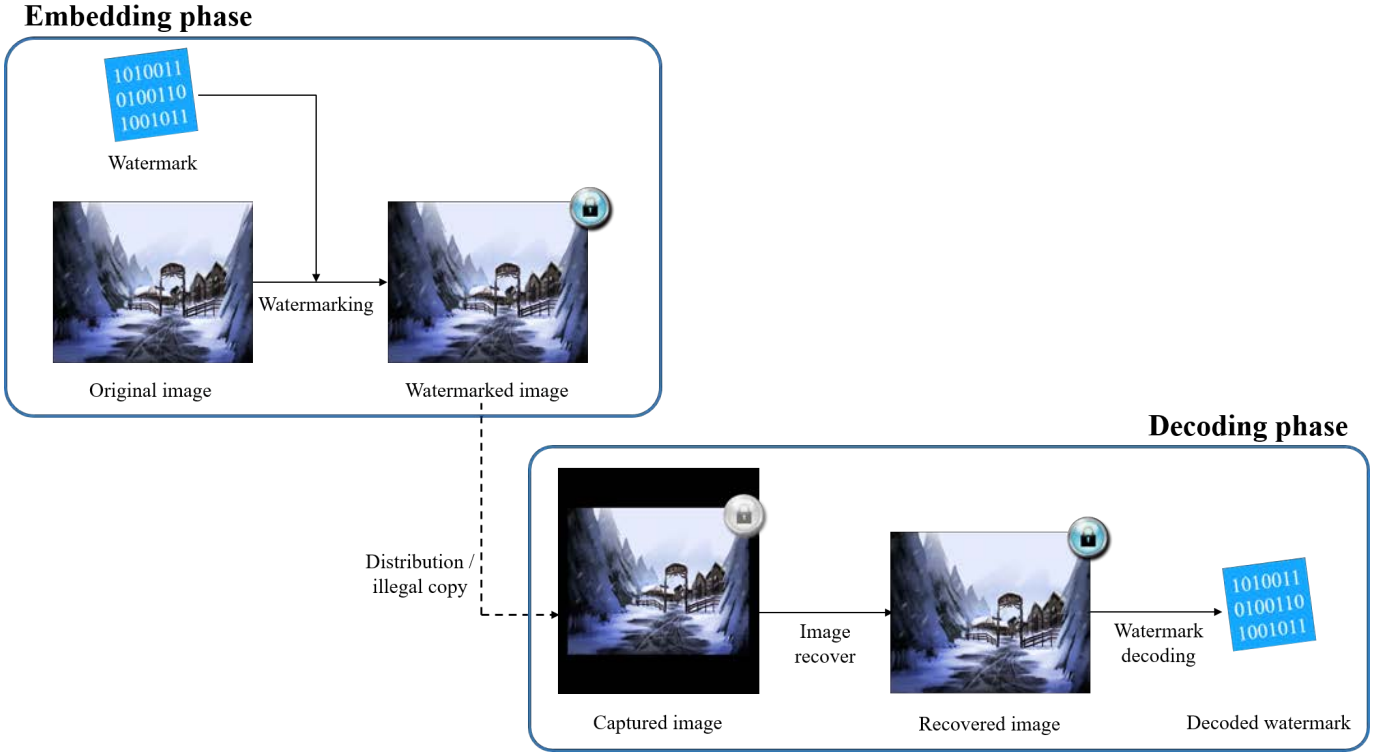

Fig. 1: Illegal distribution scenario for real-time web-comics and watermark extraction process suitable for this scenario. Illegal distribution is frequently caused by screen capturing on Internet browsers and smartphone applications. Scaling and translation occur in this process and these distortions must be corrected before watermark decoding.

form. However, this approach requires a process for finding the original image information, comparing the original image with the geometrically distorted image, and then recovering the geometrical characteristics. This process is often inefficient because it is done manually or using heuristic search.

As mentioned above, in real-time content provided by Internet browsers and applications, geometric distortion occurs very frequently due to capturing. Figure 1 shows an illegal distribution scenario that occurs in a real-time content service and a watermark extraction process suitable for such a scenario. The piracy process, such as the capturing shown in this figure, makes it difficult to decode the watermark by breaking the synchronization of the watermark. Correction of the geometric distortion must be performed in order to extract the watermark accurately. Therefore, there is a need for a method that can effectively recover the image after such geometric distortion.

In this paper, we propose a convolutional neural network (CNN)-based template that can effectively correct geometric distortion. The proposed method divides the image into blocks of a specific size, inserts bit information into half of the blocks using the conventional watermarking technique, and inserts a CNN-based template to correct the geometric distortion in the other half. By inserting these block-based templates, it is possible to compensate for the disadvantages of geometric distortion while preserving the advantages of existing watermarking techniques with the exception of the bit capacity.

Conventional template-based watermarking techniques usually utilize whole image-unit transform rather than block-unit transform. This is because it is advantageous to use the image-unit transform when searching for invariant domains and insertion methods against rotation, scaling and translation (RST) attacks. When using the block-unit transform, it is necessary to solve the problem of correcting the block synchronization after geometrical distortion.

However, the image-unit template has a problem in that the template cannot be inserted independently with the watermark containing copyright information unless they use the same transform domain. Therefore, there is a limitation that the template and watermark signal can interfere with each other. Also, it is difficult to design a multi-bit watermarking system in comparison with the block-unit method.

On the other hand, since the proposed template is a block-unit method and can be separated spatially, it does not cause interference with the watermark. In addition, because it is a learning-based method, it has the advantage of being able to respond quickly to new types of distortion. Due to these advantages, the proposed template can be applied not only to conventional watermarking methods but also to newly proposed watermarking methods for various purposes.

For example, the proposed template can be applied to new form of watermarks such as the recently proposed depth-image-based rendering (DIBR) watermarking technique. DIBR is a rendering method to give stereoscopic effect to images [19], [20], but it can not be protected by conventional watermarks and templates because it causes horizontal non-linear distortion. To cope with this, various DIBR watermarking techniques have been proposed [21]–[24], but they show weaknesses in geometric distortion since there have been few studies on the topic. The proposed learning-based and block-based template can easily solve the problem of geometric distortion for the DIBR watermark.

The insertion sequence of the proposed algorithm consists of watermark insertion (which contains bit information) and template insertion (which is used for correction of geometric distortion). The decoding sequence consists of template extraction, template matching, image recovery, and bit decoding from the watermark of the recovered image.

The template consists of a template generation network, a template extraction network, and a template matching network. Template generation and extraction networks are designed to behave similar to the insertion and extraction methods for existing template techniques. Matching networks refer to existing geometric matching networks [25]–[27].

The watermark consists of a watermark embedder and a watermark decoder. The watermark embedder and decoder use the same method as the conventional watermarking method with a domain transform, watermark insertion/decoding in the transformed domain, and inverse transform.

The contributions of the proposed method are as follows:
1) A learning-based template network that restores geometric distortion. Through learning, a robust template against various attacks can be designed.
2) Block-unit template. This makes it easy to design a multi-bit watermarking system and does not interfere with the watermark signal because it can be inserted independently with the watermark.
3) Due to the above two characteristics, the template can be easily applied to other watermarking techniques. This complements the vulnerability to geometric distortion of new forms of watermarks such as the DIBR watermarking method as well as existing watermarks.
4) Robust multi-bit watermark embedding and decoding algorithm based on the template.

The remainder of this paper is organized as follows. Section II presents the main concept of the proposed method, and Section III discusses the proposed method. Section IV presents experimental results and Section V concludes the paper.

## II. MAIN CONCEPT OF PROPOSED METHOD

The proposed method consists of a preparation step, insertion step, and decoding step as shown in Fig. 2. In the preparation step, a random binary code is generated using the key, and this code is used to generate a 2D binary template matrix $K$ of $M \times N$ size. This matrix determines whether each block is a template block or a watermark block in an image divided into blocks. It also serves as the ground truth for estimating the RST parameters in the template matching step.

In the insertion step, the image is divided into blocks, and then watermark insertion and template insertion are performed. First, the image is spatially divided into $M \times N$ blocks. The set of generated blocks is defined as $B$. The following rules distinguish the roles of the blocks.

$$B(x,y) = \begin{cases} \text{Template block, if } K(x,y) = 1 \\ \text{Watermark block, if } K(x,y) = 0 \end{cases} \quad (1)$$

where, $x$ and $y$ are the horizontal and vertical coordinates of $B$ and $K$, $0 \leq x$ ¡ $M$, and $0 \leq y$ ¡$N$.

Then, a block based watermark is inserted into all the watermark blocks through the watermark embedder. The watermark embedder consists of an image transform, watermark insertion, and inverse transform in the same manner as conventional watermarking techniques.

As the final step of the insertion, a template is inserted into the template blocks. Template insertion is completed by simply adding the $w \times h$ sized noise output from the template generation network to the image, where $w$ and $h$ are the width and height of the image. The template generation network is responsible for generating a specific form of noise that can be detected in the template extraction network.

The decoding step consists of extracting the template, estimating the distorted geometric information of the image, recovering the image, and decoding the bits from the watermark. First, in the template extraction step, the extraction network finds the location where the template is inserted in the image. The template extraction network outputs a matrix $K_r$' with a value of 1 where the template is inserted and 0 where the watermark is inserted. Inputting the $K_r$' and the original template $K_r$, which can be obtained from the key, into the template matching network yields estimated RST parameters. $K_r$ is simply a resize of $K$, which is used to increase the resolution of the template.

The geometrically distorted image is recovered using the estimated RST parameters. This step solves the problem of block synchronization, which is the biggest problem with block-based watermarking techniques. Therefore, the block-based watermark can be extracted blindly by using the block size and position information used in the watermark insertion step.

As the last step of decoding, the watermark extraction proceeds in the order of image transform and watermark extraction in the transformed domain as in conventional techniques.

The template generation network, template extraction network, and template matching network, which require a more detailed description, are described in Section III-A and Section III-B, and the watermark embedder and decoder are described in Section III-C.

## III. PROPOSED METHOD

In this section, the description of the template and the watermark are presented separately without considering the insertion and extraction order. The insertion and extraction order is detailed in Section II.

In each step of template insertion and extraction, a resized $K$ is used. Resizing is performed with the nearest neighbor filter, and $K$ resized into $r \times r$ is defined as $K_r$. For example, $K_{64}$ means that the $K$ of size $M \times N$ is resized to 64×64.

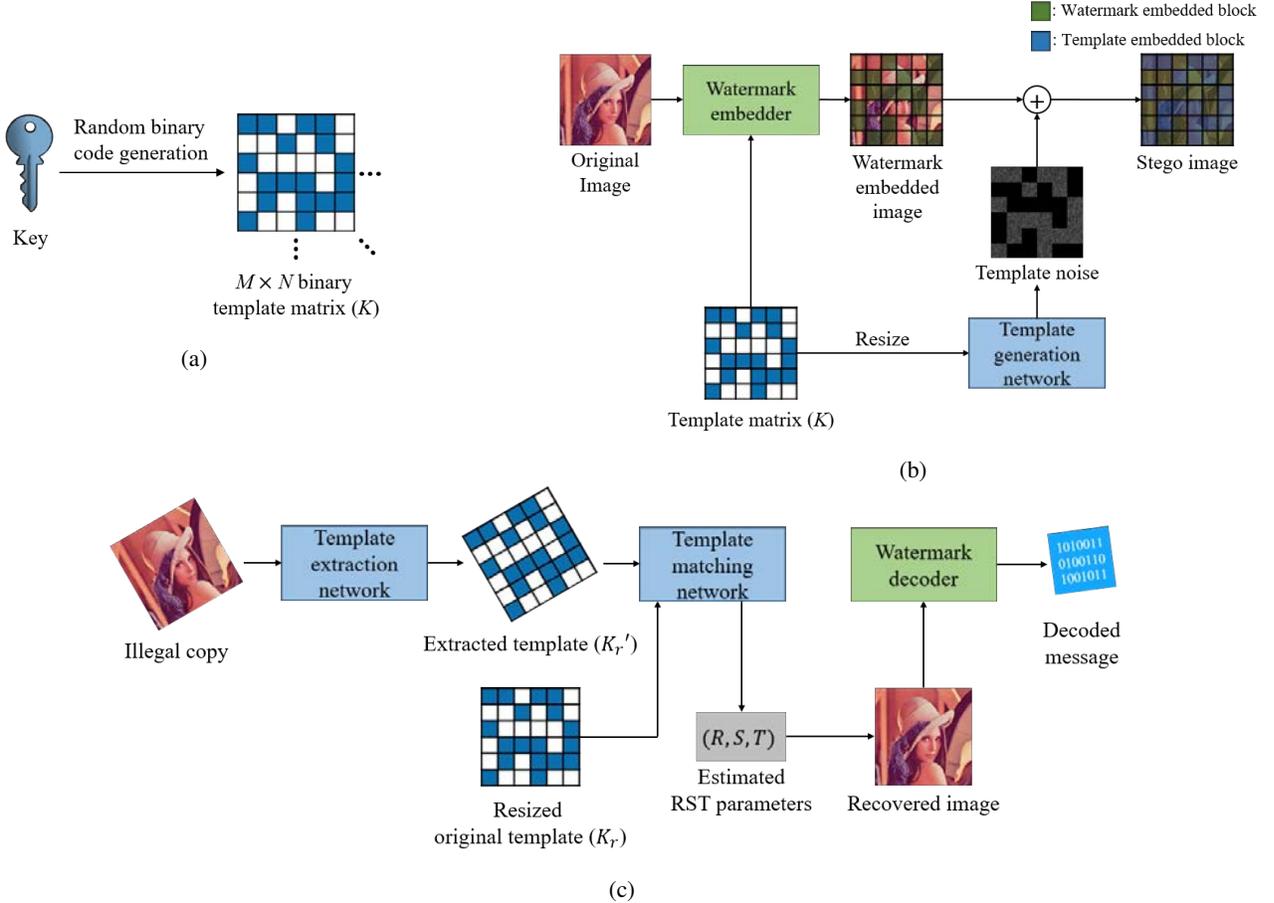

Fig. 2: An overview of the proposed method. (a) preparation step, (b) insertion step, and (c) decoding step. A Stego image is an image in which both a watermark and a template are inserted.

The reason for resizing $K$ is to increase the resolution of the template to increase the accuracy in the matching step.

### A. Template embedding and extraction network

Fig. 3 presents a template insertion and extraction scheme. Template embedding is performed as a simple sum as,

$$I_T = I + T_n, \qquad (2)$$

where $I_T$ is the template inserted image, $I$ is the original image, and $T_n$ is the template noise generated from the template generation network.

In the template extraction step, a template is extracted from $I_T$. The goal is to learn the template generation network and extraction network so that the extracted template $K'_{64}$ becomes the same as the template $K_{64}$ used at the template insertion step.

The detailed template generation and extraction network structures are shown in Fig. 4. We set the kernel size of all convolutional layers in the network to $(3, 3)$. The reason is to extract the local features of the image similar to discrete wavelet transform rather than global features similar to DFT. In order to estimate the RST parameters, the same geometric transformation as that occurring in the image must occur in the extracted template. If the template is perfectly invariant and the template is extracted without coordinate distortion, the RST parameters cannot be estimated. Thus, by exploiting local features, we induce the template to have semi-invariant properties. A semi-invariant template preserves its value from geometric distortion, but the coordinates are transformed according to the degree of geometric distortion.

The template generation network is an inverse process to the template extraction network. This template generation network generates a template noise $(T_n)$ of size $512 \times 512$ from a template $(K_r)$ of size $64 \times 64$ using a transposed convolutional layer. The process of expanding the dimension is similar to the insertion method for conventional watermarking techniques. In conventional watermarking techniques, when the watermark inserted in the middle frequency of the transformed domain passing through the inverse transform, the watermark signal spreads to the entire image of the spatial domain. The proposed template generation network also spreads the low dimensional signal to the high dimensional spatial domain similar to the conventional technique. At the end of the network, the generated template noise is masked so that the template noise does not interfere with the watermark block. The loss of the template generation network is defined as,

$$L_g = \frac{\sum_{i,j=1}^{U}(T_n^o(i,j))^2}{U^2},$$  (3)

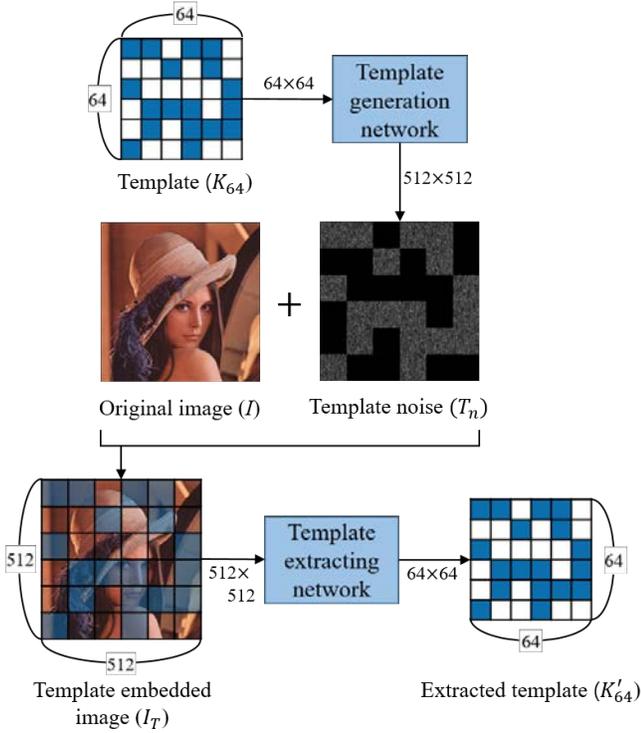

Fig. 3: Template insertion and extraction overview. Networks are trained so that the extracted template $K'_{64}$ and the inserted template $K_{64}$ have the same shape.

$T_n^o$ is the template noise generated from the generation network, $i$ and $j$ denote the horizontal and vertical coordinates of the generated template noise, respectively, $U$ represents the width and height of the network output and is set to 512. That is, since $L_g$ corresponds to the average energy of the template to be inserted, the generation network is trained so as to improve the template invisibility. $L_g$ is combined with the loss of the extraction network described below, and the generation network and the extraction network are trained together.

As in (2), simply adding the template noise generated from the template generation network to the original image will complete the template embedding. The template embedded image is then sent to the template extraction network after RST distortion.

The template extraction network extracts 64×64 templates that are the same size as the input of the generation network from the attacked image. The loss of the template extraction network is defined as,

$$L_e = \frac{1}{V^2} \sum_{i,j=1}^{V} [K'_V(i,j) - G^T(K_V(i,j))]^2,$$  (4)

where $V$ is the size of the output of the extraction network and is set to 64. $K'_V$ denotes the output of the network, i.e., the extracted template, and $K_V$ denotes the original template. $i$ and $j$ denote the horizontal and vertical coordinates of the template, respectively. $G^T$ indicates a geometric transformation using the ground-truth parameters. Since the geometric distortion that occurs in the image occurs equally in the inserted template, we define the loss function so that the extracted template is also subjected to this geometric distortion. This loss, $L_e$, trains the extraction network to improve the extraction accuracy of the template.

The total loss using the losses of the template generation and extraction networks is defined as,

$$L_t = \lambda L_g + (1 - \lambda)L_e,$$  (5)

where $\lambda$ is the trade-off parameter between invisibility and template extraction accuracy. The larger the $\lambda$, the higher the invisibility but the lower the extraction accuracy.

In order to utilize the middle frequency as the template noise similar to the existing watermarking technique, the template generation network is pre-trained before the whole network training. First, as in the spread-spectrum watermark [1], we generate a middle frequency noise with a size of 512×512. This noise is generated by substituting a pseudo-random sequence from 1/4 to 3/4 of the zigzag-scanned DCT coefficients, and is defined as $T_m$. The pseudo-random sequence is set with an average value of 0 and a variance of 1. Then the template generation network is pre-trained until $\sum_{i,j=1}^{U}(T_n^o(i,j) - T_m(i,j))^2/U^2$ is less than 0.1. The template extraction network is initialized by the Xavier uniform initializer [28].

### B. Template matching network

The template matching network shown in Fig. 5 compares the extracted template with the original template and estimates the RST parameters. First, the feature extraction network extracts features from the extracted template and the original template. The feature extraction network mimics domain transformation methods such as DFT. Domain transforms, such as DFT, are the sum of all pixels multiplied by different weights in the image. Similar to DFT, we use the kernel size as the template size to compute the global features of the template. We compute 256 global features and reshape it by 16×16. This is similar to calculating 16×16 DFT coefficients from an image. As we can estimate the translation degree from the phase of the DFT, these global features will facilitate RST parameter estimation.

The extracted features are then matched to estimate the RST parameters. These matching layers refer to the structure in [25]–[27]. Feature extraction networks are siamese networks that share weights with each other. After the feature extraction network, a concatenation layer, which is fast and has good matching accuracy, is used to combine features extracted from the feature extraction networks. Later layers are identical to the structure in [27]. The final result is a five-dimensional matrix $[R, S_x, S_y, T_x, T_y]$, which indicates the rotation, scaling of $x$ and $y$, and the translation of $x$ and $y$ parameters. The loss of the matching network is defined as,

$$L_d = \frac{1}{S^2} \sum_{i,j=1}^{S} d[G^e(x_i, y_j), G^T(x_i, y_j)]^2,$$  (6)

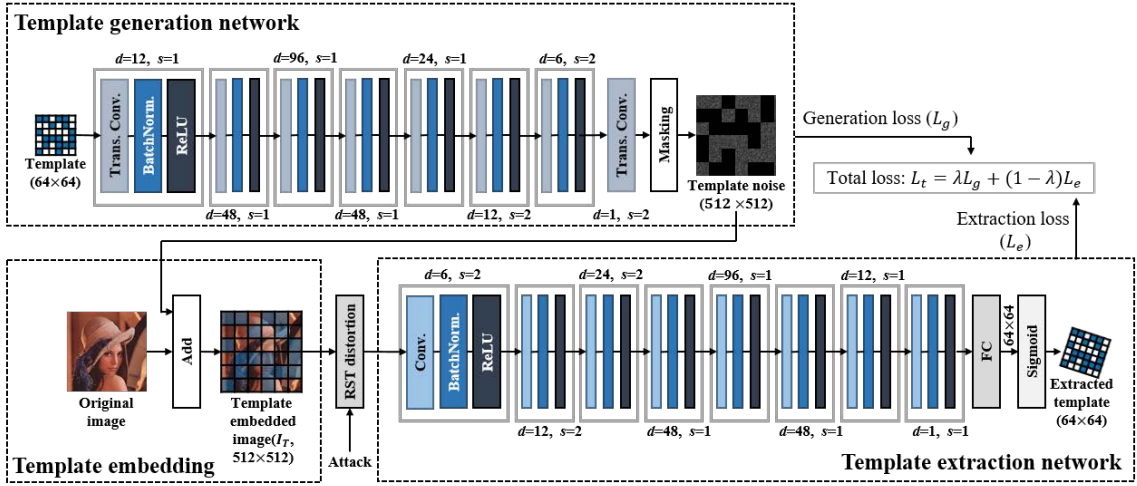

Fig. 4: Proposed template generation and extraction network architectures. All kernel sizes of convolutional layers are set to $(3, 3)$, $d$ and $s$ denote the depth and stride, respectively.

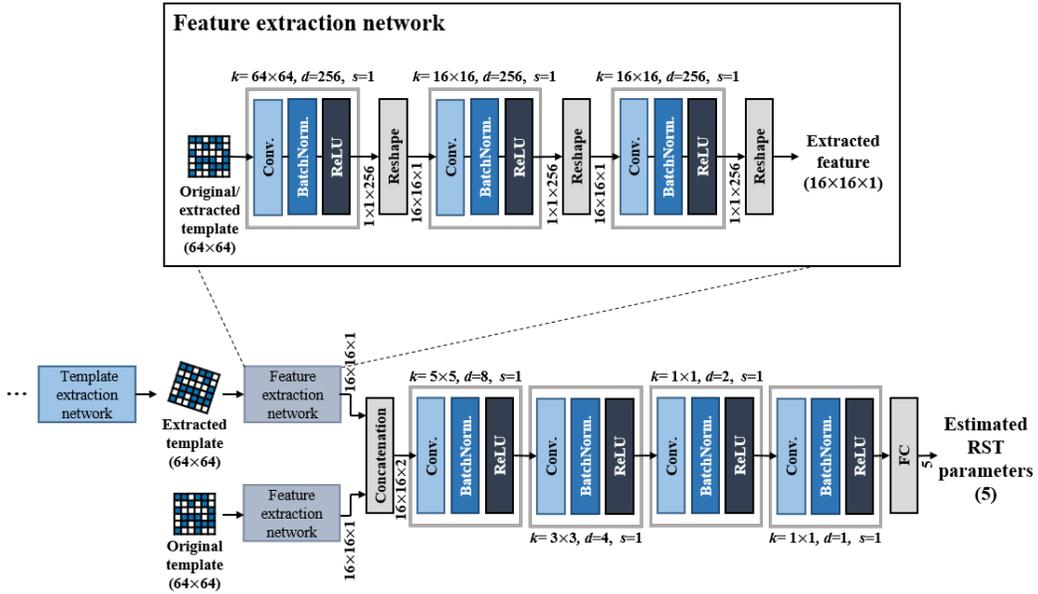

Fig. 5: Template matching network architecture. Two feature extraction networks share weights with each other. $k$ means kernel size, and zero-padding is not used in matching networks.

where $d$ is the Euclidean distance between two points, $G_e$ is the geometric transformation using the estimated RST parameters, and $G^T$ is the geometric transformation using the ground-truth RST parameters. $x_i$ and $y_i$ correspond to the $S$ equally divided points of the image in the horizontal and vertical directions, respectively, and $S$ is set to 10. That is, $L_d$ corresponds to the mean squared error between the estimated points and the true points. The reason for using the Euclidean distance as a loss without using the RST parameters directly is that each RST parameter has different weights of distortion on the image. For example, when a parameter of the same value is used, the distortion of rotation is larger than the distortion of translation.

Although the template matching network has been described separately here, the template generation and extraction net-work described above are attached to the template matching network, and then end-to-end learning is performed. End-to-end loss is defined as,

$$L_{end-to-end} = \lambda L_g + (1 - \lambda) L_d, \tag{7}$$

where $\lambda$ is the trade-off parameter between invisibility and matching accuracy. The template matching network also uses a Xavier uniform initializer to initialize the network. Since $L_d$ includes training of the template extraction network, the template generation network, template extraction network, and template matching network are trained all at once by $L_{end-to-end}$.

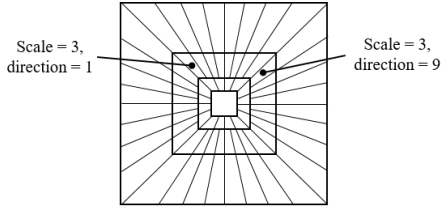

Fig. 6: Frequency spectrum coverage of curvelet transform. The curvelet decomposes the frequency domain into various scales and directions.

### C. Watermark embeder and decoder

The watermark has the role of inserting and decoding bit information. Similar to conventional techniques, the watermark is inserted/decoded in the transformed domain, and the curvelet is used as the transform domain. The reason for using this domain is that it is easy to insert multiple bits into one block through parameter adjustment while the watermark in this domain remains invisible and robust.

The curvelet transform is a multi-scale decomposition-like wavelet transform, and the curvelet represents the curve shape for various directions in the spatial domain [29]–[32]. In the image, the frequency domains are decomposed into various scales and directions by the curvelet transform as shown in Fig. 6, and the curvelet coefficients are expressed as $C^{s,l}(i, j)$, where $s$ means scale and $l$ means direction. $i$ and $j$ represent the horizontal and vertical coordinates of the coefficients in each scale and direction, respectively.

In this paper, we divide the watermark block into 5 scales and 8 directions in total. The watermark is inserted using the QIM method as in Algorithm 1 on a scale of 3 levels. In

---

**Algorithm 1** Watermark embedding procedure.

1: $A^{s,l} = \frac{1}{mn} \sum_{i=1}^{m} \sum_{j=1}^{n} abs(C^{s,l}(i, j))$
2: **if** mod[round($A^{s,l}/q$),2]=$b$ **then**
3:    $Q(A^{s,l}) = \frac{A^{s,l}}{abs(A^{s,l})} \times$ round($\frac{abs(A^{s,l})}{q}$)$\times q$
4: **else**
5:   **if** mod($\frac{abs(A^{s,l})}{q}$,1) $\leq 0.5$ **then**
6:     $Q(A^{s,l}) = \frac{A^{s,l}}{abs(A^{s,l})} \times$ round($\frac{abs(A^{s,l})}{q} + 0.5$) $\times q$
7:   **else if** mod($\frac{abs(A^{s,l})}{q}$,1) ¿ $0.5$ **then**
8:     $Q(A^{s,l}) = \frac{A^{s,l}}{abs(A^{s,l})} \times$ round($\frac{abs(A^{s,l})}{q} - 0.5$) $\times q$
9:   **end if**
10: **end if**

---

Algorithm 1, $m$ and $n$ denote the horizontal and vertical sizes of the curvelet coefficient, respectively, and $i$ and $j$ denote the horizontal and vertical coordinates of the curvelet coefficient. $Q$ is a quantization operation, $q$ is a quantization step, and $b$ is the bit to be inserted. This algorithm is the same as in other QIM techniques and is designed to find the nearest quantization level corresponding to the bit to be inserted.

Because the curvelet coefficient expresses a curved shape with directionality, it is related to the surrounding coefficients. As a result, when one coefficient is modified, the modification spreads to the surrounding coefficients [33]. Therefore, when quantization is performed, the entire $C^{s,l}(i, j)$ should be adjusted by a multiplication operation as,

$$C_m^{s,l}(i, j) = C^{s,l}(i, j) \times \frac{Q(A^{s,l})}{A^{s,l}}, \qquad (8)$$

where $C_m$ denotes the modified curvelet coefficients. $A^{s,l}$ is adjusted by applying (8) to all coefficients.

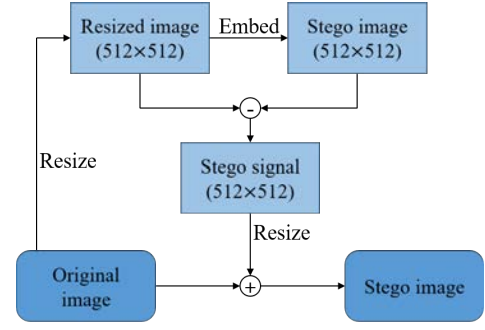

Fig. 7: The process of inserting a template and watermark into images of various sizes.

The curvelet has a symmetry property similar to DFT. Therefore, the same modification should be applied in opposite directions. Since scale 3 has a total of 16 directions, a total of 8 bits can be inserted into one block considering the symmetry property. If this process is applied to all watermark blocks, the total bit capacity becomes the number of 'watermark blocks'×8.

---

**Algorithm 2** Watermark decoding procedure.

1: $A_d^{s,l} = \frac{1}{mn} \sum_{i=1}^{m} \sum_{j=1}^{n} abs(C_d^{s,l}(i, j))$
2: $b_d = $ mod[round($A_d^{s,l}/q$),2]

---

The watermark decode proceeds as shown in Algorithm 2. $C_d$ denotes the curvelet coefficient of the image to be decoded, and $b_d$ denotes the decoded bit. This process is repeated for all watermark blocks to decode all the bits.

### D. Application method for images of various sizes

Because the input size is fixed in CNN, all descriptions are based on 512×512 images according to the network input size. In the real world, however, there are images of various sizes, so all images must be resized to 512×512 to insert and extract the template and watermark. However, if the image is resized for watermark and template insertion, image quality degradation caused by the resizing cannot be avoided.

In order to avoid image degradation due to resizing, the embedding process is performed as shown in Fig. 7. First, the image is resized to 512×512 and a template and watermark are inserted into the resized image to create a stego image. Subtracting the 512×512 image before the embedding step will leave only the stego signal, which contains the template and watermark signals. By resizing the stego signal into the original image size and adding it to the original image, the template and watermark can be inserted into the image without image quality degradation.

## IV. Experimental results

This section reports the performance of the proposed method in terms of invisibility and robustness. A comparison experiment is also conducted using the method described by Zhang [4], which uses the same domain and insertion method as the proposed method. Unlike the proposed watermarking method, Zhang's method transforms the whole image into a curvelet domain to insert a watermark. Whereas Zhang's method inserts multi-bits by dividing the curvelet directions, the proposed watermarking method inserts multi-bits by dividing the image into spatial blocks.

We also experimented with the proposed template for the new type of DIBR watermarking [22]. This DIBR watermarking method exploits the DT-CWT domain, which is robust against DIBR attacks and signal distortions, but vulnerable to geometric distortion. By comparing the performance before and after applying the template to the DIBR watermarking method, we show that the proposed template can compensate the vulnerability of the geometric distortion for the new watermarking method.

### A. Experiment Setting

BOSSBase [34], Middlebury [35], [36], and Microsoft Research 3D Video [37] datasets with 5000 images with resolutions ranging from $512 \times 512$ to $1800 \times 1500$ were used for the experiment. A total of 3500 images were used for training, and the remaining 1500 images were used for performance testing.

For network learning, 70000 images were synthesized from the 3500 images with geometric and signal distortions. The geometric distortion used random parameters of 0 to 90-degree rotation, 0.7 to 1.5 times scaling, and 0 to 30 percent translation. The signal distortion also randomly applied a Gaussian noise of 0 to 200 variance and a JPEG compression factor of 30 to 100 to the geometrically distorted image.

The proposed method was implemented with a tensorflow library [38] and Python. The network was trained using the adam optimizer [39], with a learning rate of $10^{-3}$, epsilon of $10^{-8}$, batch size of 32, and $\lambda$ in (7) of 0.2. Until convergence occurred, the end-to-end network was trained and convergence typically occurred after 15 epochs. The training took about an hour per epoch using the Nvidia GTX 1080 single GPU.

For fair comparison, all experiments were performed on a gray channel. The bit capacity for Zhang's method and the proposed watermarking method were set to 256 bits. For the proposed watermarking method, the image was divided by $8 \times 8$ to create 64 blocks. Among these, 32 blocks were used to insert bit information and the remaining blocks were used for the template. The proposed watermark was inserted on scale 3 in the curvelet domain divided by scale 5 and direction 8. Zhang's method inserts the watermark on scales 3 and 4 in the curvelet domain divided by scale 5 and direction 128. The quantization step $q$ for the proposed method and Zhang's method was set to 3.

In the experiment with the DIBR watermarking method [22], all other parameter values were set to default, and only the block size was adjusted. In the DIBR watermarking method with the template, the image was divided by $8 \times 8$ to create 64 blocks. Among these, 32 blocks were used to insert bit information and the rest were used for the template. In the method without a template, the image was divided by $5 \times 6$ to create a total of 30 blocks, and all 30 blocks were inserted with bit information. With this block size adjustment, the bit capacity of the DT-CWT method with a template and the DT-CWT method without a template were set to a similar level.

TABLE I: Average PSNR and SSIM

|  | PSNR | SSIM |
|---|---|---|
| Proposed template + watermark | 43.6668 | 0.9859 |
| Zhang's method | 43.2523 | 0.9850 |
| DT-CWT (DIBR watermark) | 40.3532 | 0.9788 |
| DT-CWT + proposed template | 41.1286 | 0.9804 |

### B. Image quality

Fig. 8 shows the original image and the template/watermark-embedded image. As can be seen, the degradation of quality due to the proposed watermark and template insertion is hardly noticeable. We also measured the peak signal-to-noise ratio (PSNR) and structural similarity (SSIM) [40] for more objective image quality measurements. As can be seen from Table I, the 'proposed template + proposed watermark' shows a similar image quality to that obtained using Zhang's method.

The watermarking method using DT-CWT shows a lower visual quality than other methods because the DT-CWT watermarking method greatly modifies the image to have robustness to DIBR. 'DT-CWT + proposed template' has a slightly better visual quality than the DT-CWT only method because it uses half the image area as a template, which has relatively lower energy than the DIBR watermark.

### C. Robustness test for RST attack

Fig. 9 shows the robustness test results for rotation, scaling, and translation. The following four methods were tested for robustness: (1) proposed watermarking method without recovery, (2) proposed watermarking method with proposed template recovery, (3) proposed watermarking method with ground-truth recovery and (4) Zhang's method. The robustness performance was measured based on the bit-error-rate (BER).

Zhang's method shows good robustness to low-level geometric distortion. This is because the absolute values of curvelet coefficients do not have a large variation in weak geometric distortions. Especially, this method shows low BER for weak scaling and translation. However, it shows a relative weakness in rotation.

The proposed watermarking method without recovery showed lower performance despite using a similar insertion method in the same domain as Zhang's method. This is because the watermark is inserted with block-units unlike with Zhang's method. Since block synchronization is broken due to geometric distortion, block-based watermarking is not effective without correction for geometric distortion. On the other hand, low BER is obtained for scaling even without recovery because the proposed watermarking method uses resized images in fixed sizes of 512×512 for the watermark insertion/decoding process to fit the image size to the network size.

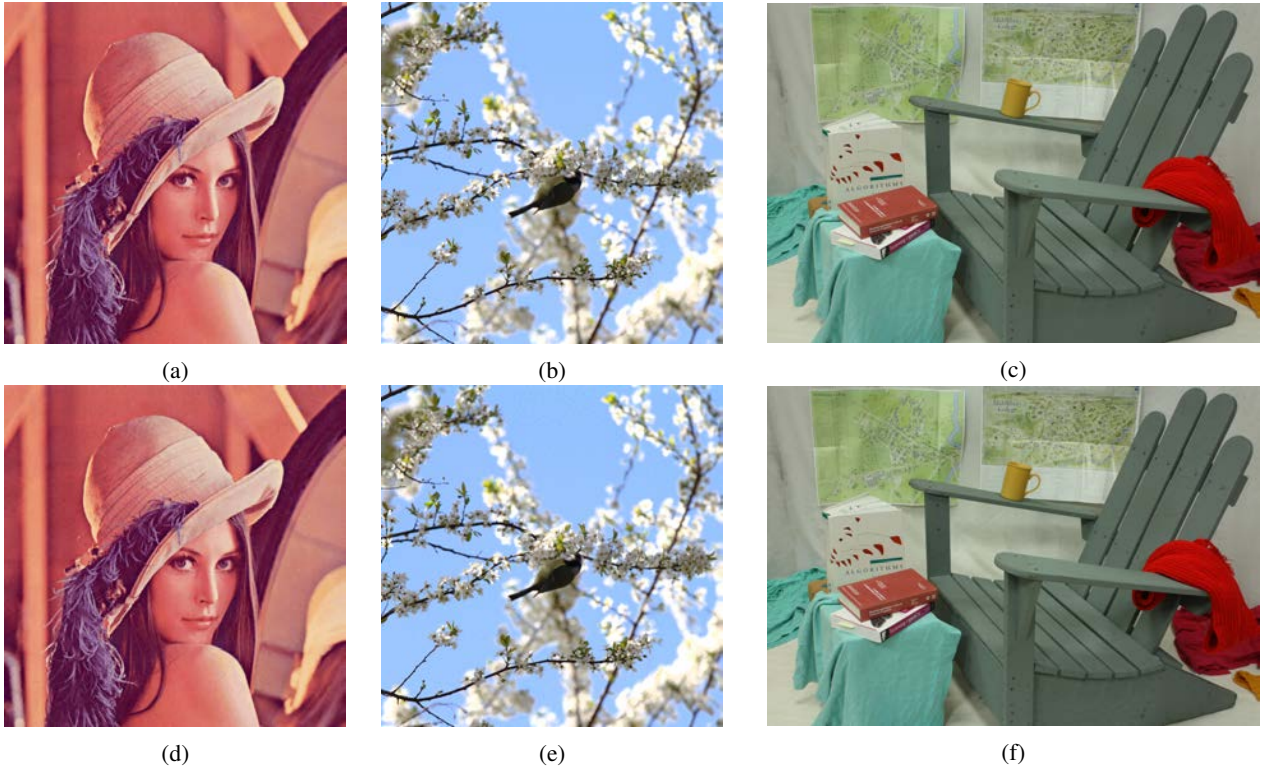

Fig. 8: (a), (b), (c) Original image, (d), (e), (f) Proposed template and watermark-embedded image.

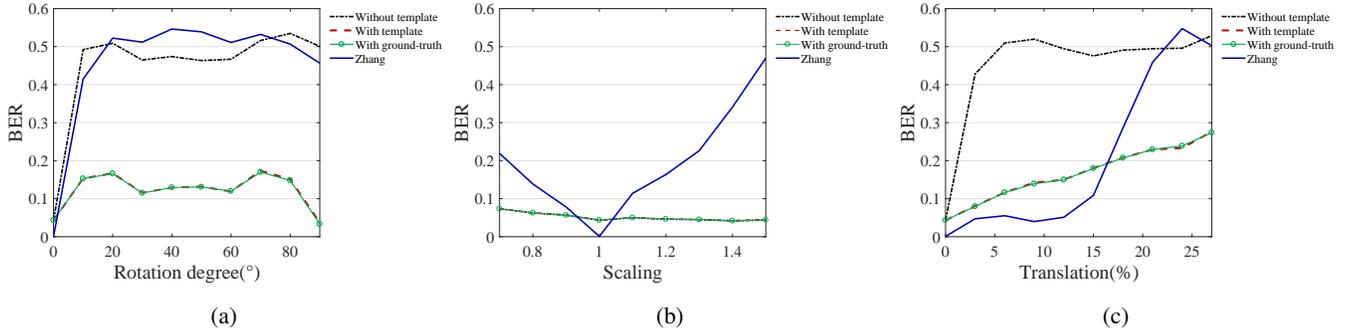

Fig. 9: Robustness results for geometric distortions. (a) rotation, (b) scaling, (c) translation. The method with the proposed template recovery and the method with ground truth recovery show almost the same performance.

If an image is recovered using the proposed template, the proposed watermark shows a low error even for strong geometric distortion. This result shows that the robustness is almost identical to the results of recovery with ground-truth, which means that the template almost completely recovers the image from the geometric distortion. The error that occurs even when the image is completely recovered is because the watermark information is cropped together with the image information. Except for the cropped part, the watermark is normally decoded.

### D. Robustness test for simultaneous attack

To test the robustness of the template for signal distortion, we measured the BER when signal distortion and RST distortion occurred simultaneously. We compared the BER when recovering an image with the proposed template and ground-truth.

Tables II, III, and IV show the results of BER when Gaussian noise and RST distortion occur simultaneously. As can be seen from these tables, the larger the noise variance, the greater the error in the method with a template compared to the method with ground-truth. At 200 noise variance, the BER is very high in the method with a template. This is because the noise has corrupted the template as well as the watermark signal, and the image has not been correctly recovered. However, considering that the variance of the inserted template

TABLE II: Average BER for rotation attack with Gaussian noise. GT denotes ground-truth.

| Rot. | Recover | Noise variance | | | |
|------|---------|----|----|----|----|
| | | 25 | 50 | 100 | 200 |
| 10° | With template | 0.174 | 0.233 | 0.268 | 0.304 |
| | With GT | 0.17 | 0.22 | 0.265 | 0.281 |
| 30° | With template | 0.138 | 0.205 | 0.258 | 0.288 |
| | With GT | 0.132 | 0.189 | 0.223 | 0.244 |
| 50° | With template | 0.169 | 0.241 | 0.28 | 0.314 |
| | With GT | 0.145 | 0.2 | 0.232 | 0.25 |
| 70° | With template | 0.204 | 0.297 | 0.333 | 0.387 |
| | With GT | 0.189 | 0.248 | 0.296 | 0.32 |
| 90° | With template | 0.081 | 0.204 | 0.254 | 0.33 |
| | With GT | 0.054 | 0.159 | 0.194 | 0.243 |

TABLE III: Average BER for scaling attack with Gaussian noise.

| Scaling | Recover | Noise variance | | | |
|---------|---------|----|----|----|----|
| | | 25 | 50 | 100 | 200 |
| 0.7 | With template | 0.089 | 0.173 | 0.235 | 0.275 |
| | With GT | 0.089 | 0.168 | 0.226 | 0.264 |
| 0.9 | With template | 0.07 | 0.16 | 0.201 | 0.249 |
| | With GT | 0.061 | 0.163 | 0.197 | 0.244 |
| 1.2 | With template | 0.068 | 0.161 | 0.207 | 0.241 |
| | With GT | 0.068 | 0.161 | 0.207 | 0.241 |
| 1.5 | With template | 0.065 | 0.152 | 0.204 | 0.243 |
| | With GT | 0.065 | 0.152 | 0.204 | 0.243 |

noise is less than 10, the noise variance of 200 is very strong, and in practice, such a large amount of noise barely occurs. In addition, there is little difference between the template-recovered method and the ground-truth-recovered method for scaling. This is because the watermark is inserted/decoded at a fixed image size of 512×512 as mentioned above.

Tables V, VI, and VII are the results of BER when JPEG compression and an RST attack occur simultaneously. These results tend to be similar to those for Gaussian noise. However, as can be seen from the results of the ground-truth-recovered

TABLE IV: Average BER for translation attack with Gaussian noise.

| Trans. | Recover | Noise variance | | | |
|--------|---------|----|----|----|----|
| | | 25 | 50 | 100 | 200 |
| 3% | With template | 0.119 | 0.229 | 0.273 | 0.338 |
| | With GT | 0.1 | 0.198 | 0.23 | 0.275 |
| 9% | With template | 0.163 | 0.249 | 0.289 | 0.354 |
| | With GT | 0.15 | 0.227 | 0.253 | 0.301 |
| 15% | With template | 0.203 | 0.291 | 0.329 | 0.387 |
| | With GT | 0.19 | 0.263 | 0.289 | 0.324 |
| 21% | With template | 0.255 | 0.328 | 0.364 | 0.427 |
| | With GT | 0.234 | 0.294 | 0.319 | 0.353 |
| 27% | With template | 0.304 | 0.369 | 0.4 | 0.445 |
| | With GT | 0.284 | 0.338 | 0.359 | 0.378 |

TABLE V: Average BER for rotation attack with JPEG.

| Rotation | Recover | JPEG quality | | |
|----------|---------|-----|----|----|
| | | 100 | 70 | 30 |
| 10° | With template | 0.153 | 0.183 | 0.218 |
| | With GT | 0.128 | 0.132 | 0.146 |
| 30° | With template | 0.122 | 0.143 | 0.199 |
| | With GT | 0.093 | 0.097 | 0.106 |
| 50° | With template | 0.157 | 0.17 | 0.216 |
| | With GT | 0.107 | 0.111 | 0.121 |
| 70° | With template | 0.191 | 0.288 | 0.31 |
| | With GT | 0.142 | 0.144 | 0.157 |
| 90° | With template | 0.08 | 0.134 | 0.206 |
| | With GT | 0.04 | 0.061 | 0.086 |

TABLE VI: Average BER for scaling attack with JPEG.

| Scaling | Recover | JPEG quality | | |
|---------|---------|-----|----|----|
| | | 100 | 70 | 30 |
| 0.7 | With template | 0.1 | 0.127 | 0.17 |
| | With GT | 0.1 | 0.127 | 0.17 |
| 0.9 | With template | 0.063 | 0.108 | 0.146 |
| | With GT | 0.063 | 0.105 | 0.146 |
| 1.2 | With template | 0.047 | 0.093 | 0.127 |
| | With GT | 0.047 | 0.090 | 0.127 |
| 1.5 | With template | 0.048 | 0.086 | 0.129 |
| | With GT | 0.044 | 0.086 | 0.129 |

method, the proposed watermarking method is robust against JPEG compression, so the overall error is lower than that for the Gaussian noise attack.

### E. Application of proposed template to DIBR watermarking method

We tested the robustness of the DIBR watermarking method against RST when the proposed template is applied. The experiment was conducted in three ways: (1) DT-CWT with template recovery, (2) DT-CWT with ground-truth recovery, and (3) Only DT-CWT method. All experiments used the right-view image rendered by DIBR. In other words, we measured the BER of images with DIBR distortion and RST distortion

TABLE VII: Average BER for translation attack with JPEG.

| Translation | Recover | JPEG quality | | |
|-------------|---------|-----|----|----|
| | | 100 | 70 | 30 |
| 3% | With template | 0.081 | 0.086 | 0.108 |
| | With GT | 0.053 | 0.056 | 0.067 |
| 9% | With template | 0.141 | 0.142 | 0.154 |
| | With GT | 0.114 | 0.115 | 0.121 |
| 15% | With template | 0.183 | 0.212 | 0.239 |
| | With GT | 0.154 | 0.156 | 0.163 |
| 21% | With template | 0.233 | 0.236 | 0.249 |
| | With GT | 0.21 | 0.209 | 0.211 |
| 27% | With template | 0.278 | 0.3 | 0.309 |
| | With GT | 0.252 | 0.253 | 0.257 |

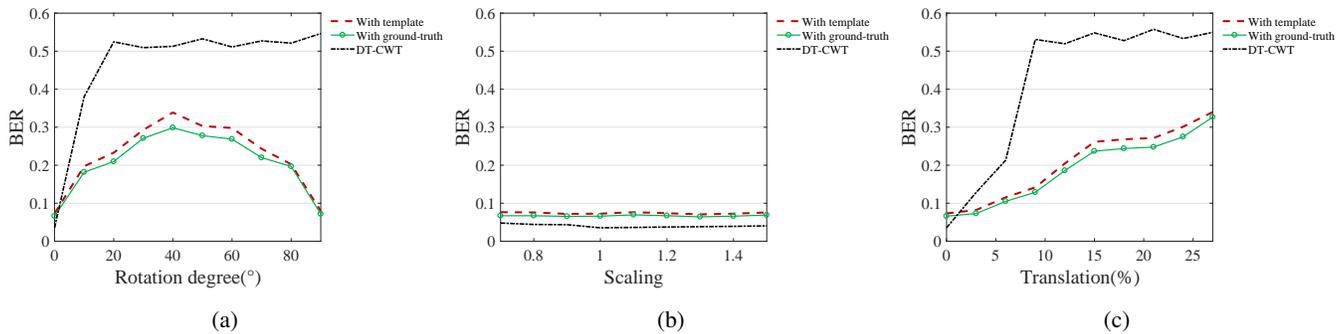

Fig. 10: Average BER for RST attack with DIBR rendering. (a) rotation, (b) scaling, (c) translation.

simultaneously. DIBR parameters used in this test were the recommended values in [41].

As shown in Fig. 10, the DT-CWT-based DIBR watermarking method has a low BER for weak geometric distortions, similar to curvelet watermarking methods. However, as the degree of geometric distortion increases, the BER increases sharply.

On the other hand, if the image is recovered using a template with the DT-CWT method, the watermark can be decoded well after the geometric distortion. The DT-CWT method recovered with a template has a slightly higher BER than that recovered with ground-truth because the template is disturbed by DIBR. However, since the error increase rate is insignificant, the proposed template can be considered robust to the new type of distortion, DIBR. These results show that the proposed template can give robustness against geometric attacks to newly proposed watermarking techniques.

## V. Conclusion

As the market for real-time contents continues to increase, the problem of illegal copying of contents is also increasing. These real-time contents are often geometrically distorted by capturing, and many existing watermarking methods are vulnerable to geometric distortion. In this paper, we proposed a CNN-based template that recovers images from geometric distortion. This template is inserted spatially separated from the watermark containing the copyright information, allowing correction of geometric distortions with a small loss of watermark performance and blind detection of the watermark. In addition, the proposed template can be applied to new watermarking techniques for various types of contents such as 3D images. Especially, the DIBR 3D content watermarking method has not been studied much yet, so the problem of vulnerability to geometric distortion has not been solved. Experimental results show that the proposed learning-based template can compensate for the vulnerability of this new type of watermarking technique to geometric distortion as well as typical 2D image watermarking techniques. Watermarking techniques and templates using CNN are still in the early stage. Further research is needed to improve the watermark performance such as masking techniques that will increase invisibility and security enhancement. In addition, we will extend the scope of our research into video content watermarking using CNN.

## References

[1] M. Barni, F. Bartolini, V. Cappellini, and A. Piva, "A dct-domain system for robust image watermarking," *Signal processing*, vol. 66, no. 3, pp. 357–372, 1998.

[2] V. Solachidis and L. Pitas, "Circularly symmetric watermark embedding in 2-d dft domain," *IEEE transactions on image processing*, vol. 10, no. 11, pp. 1741–1753, 2001.

[3] D. Simitopoulos, D. E. Koutsonanos, and M. G. Strintzis, "Robust image watermarking based on generalized radon transformations," *IEEE Transactions on Circuits and Systems for Video Technology*, vol. 13, no. 8, pp. 732–745, 2003.

[4] C. Zhang, L. L. Cheng, Z. Qiu, and L.-M. Cheng, "Multipurpose watermarking based on multiscale curvelet transform," *IEEE Transactions on Information Forensics and Security*, vol. 3, no. 4, pp. 611–619, 2008.

[5] L. E. Coria, M. R. Pickering, P. Nasiopoulos, and R. K. Ward, "A video watermarking scheme based on the dual-tree complex wavelet transform," *IEEE Transactions on Information Forensics and Security*, vol. 3, no. 3, pp. 466–474, 2008.

[6] M. A. Akhaee, S. M. E. Sahraeian, and F. Marvasti, "Contourlet-based image watermarking using optimum detector in a noisy environment," *IEEE Transactions on Image Processing*, vol. 19, no. 4, pp. 967–980, 2010.

[7] I. J. Cox, J. Kilian, F. T. Leighton, and T. Shamoon, "Secure spread spectrum watermarking for multimedia," *IEEE transactions on image processing*, vol. 6, no. 12, pp. 1673–1687, 1997.

[8] B. Chen and G. W. Wornell, "Quantization index modulation: A class of provably good methods for digital watermarking and information embedding," *IEEE Transactions on Information Theory*, vol. 47, no. 4, pp. 1423–1443, 2001.

[9] F. Ourique, V. Licks, R. Jordan, and F. Pérez-González, "Angle qim: A novel watermark embedding scheme robust against amplitude scaling distortions," in *Acoustics, Speech, and Signal Processing, 2005. Proceedings.(ICASSP'05). IEEE International Conference on*, vol. 2. IEEE, 2005, pp. ii–797.

[10] E. Nezhadarya, Z. J. Wang, and R. K. Ward, "Robust image watermarking based on multiscale gradient direction quantization," *IEEE Transactions on Information Forensics and Security*, vol. 6, no. 4, pp. 1200–1213, 2011.

[11] J. J. O. Ruanaidh and T. Pun, "Rotation, scale and translation invariant spread spectrum digital image watermarking1," *Signal processing*, vol. 66, no. 3, pp. 303–317, 1998.

[12] H. S. Kim and H.-K. Lee, "Invariant image watermark using zernike moments," *IEEE transactions on Circuits and Systems for Video Technology*, vol. 13, no. 8, pp. 766–775, 2003.

[13] Y. Xin, S. Liao, and M. Pawlak, "A multibit geometrically robust image watermark based on zernike moments," in *Pattern Recognition, 2004. ICPR 2004. Proceedings of the 17th International Conference on*, vol. 4. IEEE, 2004, pp. 861–864.

[14] S. Wang, C. Cui, and X. Niu, "Watermarking for dibr 3d images based on sift feature points," *Measurement*, vol. 48, pp. 54–62, 2014.

[15] S. Stankovic, I. Djurovic, and I. Pitas, "Watermarking in the space/spatial-frequency domain using two-dimensional radon-wigner distribution," *IEEE transactions on Image Processing*, vol. 10, no. 4, pp. 650–658, 2001.

[16] W. Lu, H. Lu, and F.-L. Chung, "Feature based watermarking using watermark template match," *Applied Mathematics and Computation*, vol. 177, no. 1, pp. 377–386, 2006.

[17] P. Bas, J.-M. Chassery, and B. Macq, "Geometrically invariant watermarking using feature points," *IEEE transactions on image Processing*, vol. 11, no. 9, pp. 1014–1028, 2002.

[18] S. Pereira and T. Pun, "Robust template matching for affine resistant image watermarks," *IEEE transactions on image Processing*, vol. 9, no. 6, pp. 1123–1129, 2000.

[19] C. Fehn, P. Kauff, M. O. De Beeck, F. Ernst, W. Ijsselsteijn, M. Pollefeys, L. Van Gool, E. Ofek, and I. Sexton, "An evolutionary and optimised approach on 3d-tv," in *Proc. of IBC*, vol. 2, 2002, pp. 357–365.

[20] C. Fehn, "Depth-image-based rendering (dibr), compression, and transmission for a new approach on 3d-tv," in *Stereoscopic Displays and Virtual Reality Systems XI*, vol. 5291. International Society for Optics and Photonics, 2004, pp. 93–105.

[21] Y.-H. Lin and J.-L. Wu, "A digital blind watermarking for depth-image-based rendering 3d images," *IEEE transactions on Broadcasting*, vol. 57, no. 2, pp. 602–611, 2011.

[22] H.-D. Kim, J.-W. Lee, T.-W. Oh, and H.-K. Lee, "Robust dt-cwt watermarking for dibr 3d images," *IEEE Transactions on Broadcasting*, vol. 58, no. 4, pp. 533–543, 2012.

[23] S.-H. Nam, W.-H. Kim, S.-M. Mun, J.-U. Hou, S. Choi, and H.-K. Lee, "A sift features based blind watermarking for dibr 3d images," *Multimedia Tools and Applications*, pp. 1–40, 2017.

[24] A. Al-Haj, M. E. Farfoura, and A. Mohammad, "Transform-based watermarking of 3d depth-image-based-rendering images," *Measurement*, vol. 95, pp. 405–417, 2017.

[25] D. DeTone, T. Malisiewicz, and A. Rabinovich, "Deep image homography estimation," *arXiv preprint arXiv:1606.03798*, 2016.

[26] A. Kanazawa, D. W. Jacobs, and M. Chandraker, "Warpnet: Weakly supervised matching for single-view reconstruction," in *Proceedings of the IEEE Conference on Computer Vision and Pattern Recognition*, 2016, pp. 3253–3261.

[27] I. Rocco, R. Arandjelovic, and J. Sivic, "Convolutional neural network architecture for geometric matching," in *Proc. CVPR*, vol. 2, 2017.

[28] X. Glorot and Y. Bengio, "Understanding the difficulty of training deep feedforward neural networks," in *Proceedings of the thirteenth international conference on artificial intelligence and statistics*, 2010, pp. 249–256.

[29] E. J. Candes and D. L. Donoho, "Curvelets: A surprisingly effective nonadaptive representation for objects with edges," Stanford Univ Ca Dept of Statistics, Tech. Rep., 2000.

[30] E. J. Candès and F. Guo, "New multiscale transforms, minimum total variation synthesis: Applications to edge-preserving image reconstruction," *Signal Processing*, vol. 82, no. 11, pp. 1519–1543, 2002.

[31] E. J. Candès and D. L. Donoho, "New tight frames of curvelets and optimal representations of objects with piecewise c2 singularities," *Communications on pure and applied mathematics*, vol. 57, no. 2, pp. 219–266, 2004.

[32] E. Candes, L. Demanet, D. Donoho, and L. Ying, "Fast discrete curvelet transforms," *Multiscale Modeling & Simulation*, vol. 5, no. 3, pp. 861–899, 2006.

[33] W.-H. Kim, S.-H. Nam, and H.-K. Lee, "Blind curvelet watermarking method for high-quality images," *Electronics Letters*, vol. 53, no. 19, pp. 1302–1304, 2017.

[34] P. Bas, T. Filler, and T. Pevnỳ, "" break our steganographic system": The ins and outs of organizing boss," in *International Workshop on Information Hiding*. Springer, 2011, pp. 59–70.

[35] H. Hirschmuller and D. Scharstein, "Evaluation of cost functions for stereo matching," in *Computer Vision and Pattern Recognition, 2007. CVPR'07. IEEE Conference on*. IEEE, 2007, pp. 1–8.

[36] D. Scharstein, H. Hirschmüller, Y. Kitajima, G. Krathwohl, N. Nešić, X. Wang, and P. Westling, "High-resolution stereo datasets with subpixel-accurate ground truth," in *German Conference on Pattern Recognition*. Springer, 2014, pp. 31–42.

[37] C. L. Zitnick, S. B. Kang, M. Uyttendaele, S. Winder, and R. Szeliski, "High-quality video view interpolation using a layered representation," in *ACM transactions on graphics (TOG)*, vol. 23, no. 3. ACM, 2004, pp. 600–608.

[38] M. Abadi, P. Barham, J. Chen, Z. Chen, A. Davis, J. Dean, M. Devin, S. Ghemawat, G. Irving, M. Isard *et al.*, "Tensorflow: A system for large-scale machine learning." in *OSDI*, vol. 16, 2016, pp. 265–283.

[39] D. P. Kingma and J. Ba, "Adam: A method for stochastic optimization," *arXiv preprint arXiv:1412.6980*, 2014.

[40] Z. Wang, A. C. Bovik, H. R. Sheikh, and E. P. Simoncelli, "Image quality assessment: from error visibility to structural similarity," *Image Processing, IEEE Transactions on*, vol. 13, no. 4, pp. 600–612, 2004.

[41] L. Zhang and W. J. Tam, "Stereoscopic image generation based on depth images for 3d tv," *IEEE Transactions on broadcasting*, vol. 51, no. 2, pp. 191–199, 2005.